\documentclass[12pt]{article}
\usepackage[english]{babel}
\usepackage{amsmath}
\usepackage{amsthm}
\usepackage{amsfonts}
\usepackage{amssymb}
\usepackage{euscript}
\usepackage[cp1251]{inputenc}
\usepackage[pdftex]{graphicx}
\newcommand{\comment}[1]{}
\usepackage[12pt]{extsizes}

\theoremstyle{definition}

\begin{document}
\let\thefootnote\relax\footnote{This research was supported by grant RSF 16-11-10260 and was done at the faculty of Mechanics and Mathematics, department of Geometry and Topology of Moscow State University}
\begin{flushleft}
\textbf{\Large{Commuting differential operators of rank 2 with rational coefficients}}
\end{flushleft}
\begin{flushleft}
\textbf{Vardan Oganesyan}
\end{flushleft}
\emph{Department of Geometry and Topology, Faculty of Mechanics and Mathematics, Lomonosov Moscow State University, Moscow, 119991 Russia.\\
E-mail address: vardan.o@mail.ru}\\
\\
\textbf{Abstract.} In this paper we find new self-adjoint commuting operators of rank 2 with rational coefficients and prove that any elliptic and hyperelliptic curves of genus 2 are spectral curves of commuting operators with rational coefficients. Also the case when curves of genus 3 are spectral curves of commuting operators with rational coefficients is studied.\\
\begin{center}
   \textbf{Introduction}
\end{center}
The commutativity condition of two differential operators
\begin{equation*}
L_n= \sum\limits^{n}_{i=0} u_i(x)\partial_x^i,  \quad  L_m= \sum\limits^{m}_{i=0} v_i(x)\partial_x^i
\end{equation*}
is equivalent to a very complicated system of non-linear differential equations. If two differential operators commute, then there is a nonzero polynomial $R(z,w)$ such that  $R(L_n,L_m)=0$ (see ~\cite{Chaundy}). The curve $\Gamma$ defined by $R(z,w)=0$ is called the \emph{spectral curve}. The genus of the curve $R(z, w) = 0$ is called the genus of commuting
pair. If\\
\begin{equation*}
L_n \psi=z\psi, \quad  L_m \psi=w\psi,
\end{equation*}
then $(z,w) \in \Gamma$. For almost all $(z,w) \in \Gamma$ the dimension of the space of common eigenfunctions $\psi$ is the same. The dimension of the space of common eigenfunctions for generic $P=(z,w) \in \Gamma$ is called the \emph{rank}. The rank is a common divisor of m and n.

Commutative rings of such operators were classified by Krichever \cite{ringkrichever}. The ring is determined by the spectral curve and some additional spectral data. If the rank equals 1, then there are explicit formulas for coefficients of commutative operators in terms of Riemann theta-functions (see ~\cite{theta}).

If the rank equals 1, then common eigenfunction $\psi(x,P)$ is Baker-Akhiezer function, where $P \in \Gamma$. Baker-Akhiezer function is a function with the following properties.\\
\\
1) Function $\psi(x,P)$  has one essential singularity at a fixed point $q \in \Gamma$
\begin{equation*}
\psi(x,P) = e^{kx}(1 + \dfrac{\xi_1(x)}{k} + \dfrac{\xi_2(x)}{k^2} +...),
\end{equation*}
where $k^{-1}$ is a local parameter in a neighborhood of $q$.\\
2) Function $\psi$  has simple poles at some points $\gamma_1, ..., \gamma_g$, where g is the genus of $\Gamma$.\\

The set $\{\Gamma, q, \gamma_1,...,\gamma_g\}$ is called the spectral data. If we take the spectral data where $D=\gamma_1 + ...+\gamma_g$ is a non-special divisor, then there is a unique function $\psi(x,P)$ satisfying the conditions 1) and 2). Let us mention that if the spectral curve is singular, then in general $\psi$ is not a function on the spectral curve, but $\psi$ is a section of a torsion-free sheaf on $\Gamma\setminus\{q\}$ (see \cite{Mumford}).

The case when rank is greater than $1$ is much more difficult. The first examples of commuting ordinary differential operators of the nontrivial rank 2 and the nontrivial genus $g=1$ were constructed by Dixmier ~\cite{Dixmier} for the nonsingular elliptic spectral curve $w^2=z^3-\alpha$, where $\alpha$ is an arbitrary nonzero constant:
\begin{equation*}
L= (\partial_x^2 + x^3 + \alpha)^2 + 2x ,
\end{equation*}
\begin{equation*}
M= (\partial_x^2 + x^3 + \alpha)^3 + 3x\partial_x^2 + 3\partial_x + 3x(x^2+\alpha).
\end{equation*}
Operators $L$ and $M$ is the commuting pair of the Dixmier operators of rank 2, genus 1.

The general classification of commuting ordinary differential operators of rank greater than 1 was obtained by Krichever ~\cite{ringkrichever}. The general form of commuting operators of rank 2 for an arbitrary elliptic spectral curve was found by Krichever and Novikov ~\cite{novkrich}. The general form of operators of rank 3 for an arbitrary elliptic spectral curve was found by Mokhov ~\cite{Mokhov1},~\cite{Mokhov2}. Moreover, examples of commuting ordinary differential operators of arbitrary genus and arbitrary rank with polynomial coefficients were constructed in  ~\cite{Mokhov4}, ~\cite{Mokhov3}.

Mironov in ~\cite{Mironov} constructed examples of commuting operators
\begin{equation*}
L = (\partial_x^2 + A_3x^3+ A_2x^2 + A_1x + A_0 )^2 + g(g+1)A_3x ,
\end{equation*}
\begin{equation*}
M^2 = L^{2g+1} + a_{2g}L^{2g} + ... + a_1L + a_0 ,
\end{equation*}
where $a_i$  are some constants and $A_i$, $A_3 \neq 0$, are arbitrary constants. Operators $L$ and $M$ are commuting operators of rank 2, genus g.

Let us suppose that function $\mathcal{P}$ satisfies the equation
\begin{equation*}
(\mathcal{P'}(x))^2 = g_2\mathcal{P}^2(x) + g_1\mathcal{P}(x) + g_0,  \quad g_2 \neq 0,
\end{equation*}
where $g_1$ and $g_2$ are arbitrary constants. In ~\cite {Mironov2} it was proved by Mironov that operators $L_1$ and $M_1$,
\begin{equation*}
L_1 = (\partial_x^2 + \alpha_1 \mathcal{P}(x) + \alpha_0)^2 + \alpha_1g_2g(g+1)\mathcal{P}(x), \quad \alpha_1 \neq 0 ,
\end{equation*}
\begin{equation*}
M_1^2 = L_1^{2g+1} + a_{2g}L_1^{2g} + ... + a_1L_1 + a_0 ,
\end{equation*}
where $a_i$  are some constants, $\alpha_i$ are arbitrary constants, is a commuting pair of rank 2, genus $g$.

Let $\wp(x)$ be the Weierstrass elliptic function satisfying the equation $\wp'(x) = 4\wp^3(x) + g_2\wp(x) + g_3$. Mironov proved in ~\cite{Mironov2} that operators $L_2$ and $M_2$,
\begin{equation*}
L_2 = (\partial_x^2 + \alpha_1 \wp(x) + \alpha_0)^2 + s_1 \wp(x) + s_2 \wp^2(x),
\end{equation*}
\begin{equation*}
M_2^2 = L_2^{2g+1} + b_{2g}L_2^{2g} + ... + b_1L_2 + b_0,
\end{equation*}
where  $b_i$ are some constants, $\alpha_1=\dfrac{1}{4} - 2g^2 - 2g$,
$s_1=\dfrac{1}{4}g(g+1)(16\alpha_0 + 5g_2)$,  $s_2=-4g(g+2)(g^2-1)$ and $\alpha_0$ is an arbitrary constant, are also a commuting pair of rank 2, genus $g$.

Examples of commuting operators of rank two with trigonometric coefficients were found in ~\cite{Davletshina}.

Let us consider
\begin{equation*}
\begin{gathered}
L_1=(\partial_x^2 + A_6x^6 + A_2x^2 )^2 + 16g(g+1)A_6x^4,\\
L_2=(\partial_x^2 + A_4x^4 + A_2x^2 + A_0)^2 + 4g(g+1)A_4x^2,
\end{gathered}
\end{equation*}
where $g \in \mathbb{N}$, $A_6 \neq 0$, $A_4\neq 0$, $A_2, A_0$ are arbitrary constants. Operators $L_1$ and $L_2$ commute with some differential operators $M_1$ and $M_2$  of order 4g+2 respectively (see~\cite{Vartan}, \cite{Vartan2}). The spectral curves of these operators have the form $w^2=z^{2m+1} + a_{2m}z^{2m} + ...+a_1z + a_0$. Common eigenfunction of operators $L_1$, $M_1$ and $L_2$, $M_2$ in some special cases were found in ~\cite{Vartan4}. Moreover, the following theorems are proved in ~\cite{Vartan}, \cite{Vartan2}.\\
\\
\textbf{1)}  If $L=(\partial_x^2 + A_nx^n + A_{n-1}x^{n-1} + ... + A_0)^2 + B_kx^{k} + B_{k-1}x^{k-1} + ...+B_0$, where $n>3, n \in \mathbb{N}$, $A_n \neq 0$, $B_k \neq 0$, commutes with a differential operator M of order $4g+2$ and M, L are operators of rank 2, then $k=n-2$ and $B_k=(n-2)^2m(m+1)A_n$ for some $m \in \mathbb{N}$.\\
\textbf{2)}For $n>6$  operator $L=(\partial_x^2 + A_nx^n)^2 + B_{n-2}x^{n-2}$ does not commute with any differential operator $M$ of order $4g+2$, where $M$ and $L$ could be a pair of rank 2.\\
\textbf{3)} If $n=5$, then $L=(\partial_x^2 + Ax^5)^2 + 18Ax^{3}$, $A\neq 0$,  commutes with a differential operator M of order $4g+2$ for all $g$ and M,L are operators of rank 2. \\
\textbf{4)} The operator  $(\partial_x^2 + Ax^5)^2 + 9m(m+1)Ax^{3}$, $A\neq 0$ does not commute with any differential operator M of order $4g+2,$ for $m>1$, where M and L could be a pair of rank 2.\\

Commuting operators $L_4$ and $M$, where $L_4 = \partial_x^4 + u(x)$ and $M^2 = L_4^{2g+1} + \beta_{2g}L_4^{2g} + ...+ \beta_0$ were studied in  ~\cite{Vartan3}.

Mironov and Zheglov (see ~\cite{Zheglov}) proved that for arbitrary integer $m$ and arbitrary spectral curve $\Gamma$ given by equation $w^2 = z^3 + c_2z^2 + c_1z + c_0$ there are polynomials
\begin{equation*}
V_m = \alpha_{m+2}x^{m+2} + ... + \alpha_0, \quad W_m = \beta_mx^m + ... + \beta_0
\end{equation*}
such that  operator
\begin{equation*}
L_{4,m} =(\partial_x^2 + V_m)^2 + W_m
\end{equation*}
commutes with a six order operator $L_{6,m}$. And the spectral curve $L_{4,m}, L_{6,m}$ coincides with $\Gamma$. Also, Mironov and Zheglov studied in ~\cite{Zheglov} automorphisms of the first Weyl algebra.\\

\textbf{Theorem 1.} \emph{The operator
\begin{equation}
L=(\partial_x^2 + A_6x^6 + A_2x^2 + \dfrac{A_{-2}}{x^2} + \dfrac{A_{-6}}{x^6} )^2 + 16g(g+1)A_6x^4 + B_0,
\end{equation}
where $g \in \mathbb{N}$, $A_6 \neq 0$ and $A_2, A_{-2}, A_{-6}, B_0$ are arbitrary constants, commutes with a differential operator $M$ of order $4g+2$. The spectral curve has the form $w^2=z^{2m+1} + a_{2m}z^{2m} + ...+a_1z + a_0$. The operators $L$ and $M$ are operators of rank 2}.\\

\textbf{Theorem 2.} \emph{The operator
\begin{equation}
L=(\partial_x^2 + A_4x^4 + A_2x^2 + A_0 + \dfrac{A_{-2}}{x^2})^2 + 4g(g+1)A_4x^2 + B_0,
\end{equation}
where  $g \in \mathbb{N}$, $A_4\neq 0$ and $A_2, A_0, A_{-2}, B_0$ are arbitrary constants, commutes with a differential operator M of order $4g+2$. The spectral curve has the form $w^2=z^{2m+1} + a_{2m}z^{2m} + ...+a_1z + a_0$. The operators $L$ and $M$ are operators of rank 2}.\\

\textbf{Theorem 3.}

\emph{1) Any elliptic curve}
\begin{equation*}
w^2 =z^3 + \beta_2z^2 + \beta_1z + \beta_0,
\end{equation*}
\emph{where $\beta_i$ are arbitrary constants, is spectral curve of commuting operators from Theorem 1.}

2) \emph{Any hyperelliptic curve of the form }
\begin{equation*}
w^2 =z^5 +\beta_4z^4 + \beta_3z^3 + \beta_2z^2 + \beta_1z + \beta_0,
\end{equation*}
\emph{where $\beta_i$ are arbitrary constants, is spectral curve of commuting operators from Theorems 1 or 2.}

3) \emph{And any hyperelliptic curve of the form }
\begin{equation*}
w^2 =z^7 + \beta_6z^6 + \beta_5z^5 +\beta_4z^4 + \beta_3z^3 + \beta_2z^2 + F_1z + F_0,
\end{equation*}
\emph{where $\beta_i$ are arbitrary constant and  $F_1=F_1(\beta_6,...,\beta_2)$ and $F_0 = \beta(\beta_6,...,\beta_2)$ are some constants, is spectral curve of commuting operators from Theorems 1 or 2}.
\\
Note that Theorem 3 doesn't say that any hyperelliptic curve of genus 2 is spectral curve of operators only from Theorem 1 or only from Theorem 2. But  any hyperelliptic curve of genus 2 is spectral curve of operators from Theorem 1 or Theorem 2.
\\
\begin{center}
\textbf{Acknowledgments}
\end{center}

The author wishes to express gratitude to Professor O. I. Mokhov for advices and help in writing this paper.
\begin{center}
   \textbf{Commuting operators of rank 2}
\end{center}

Let us consider the operator
\begin{equation}
L=(\partial^2_x + V(x))^2 + W(x)
\end{equation}
It is proved in \cite{Mironov} that the operator $L$  commutes with an operator $M$ of order $4g+2$, operators $L, M$ are operators of rank $2$ and the spectral curve of $L, M$ has the form
\begin{equation*}
w^2 = z^{2g+1} + \beta_{2g}z^{2g} + ... + \beta_0
\end{equation*}
if and only if there exists a polynomial
\begin{equation*}
Q=z^g + a_1(x)z^{g-1} + a_2(x)z^{g-2} + ...+a_{g-1}(x)z + a_g(x)
\end{equation*}
that the following relation is satisfied
\begin{equation}
Q^{(5)} + 4VQ''' + 6V'Q'' + 2Q'(2z-2W+V'') - 2QW'\equiv 0,
\end{equation}
where $Q'$ means $\partial_xQ$. The spectral curve of commuting pair $L, M$ has the form
\begin{equation}
4w^2=4F(z)=4(z-W)Q^2 - 4V(Q')^2 + (Q'')^2 - 2Q'Q''' + 2Q(2V'Q' + 4VQ'' + Q^{(4)}).
\end{equation}
Using (4), we have
\begin{equation*}
4Q'z\equiv -Q^{(5)} - 4VQ''' - 6V'Q'' - 2Q'V'' + 2QW' + 4Q'W.
\end{equation*}
We get the following system.
\begin{equation*}
\left\{
 \begin{array}{l}
a_1=W/2 + C_1\\
4a_2' = -a_1^{(5)} - 4Va_1''' - 6V'a_1'' - 2a_1'V'' + 2a_1W' + 4a_1'W\\
...\\
4a_{i+1}' = -a_i^{(5)} - 4Va_i''' - 6V'a_i'' - 2a_i'V'' + 2a_iW' + 4a_i'W\\
...\\
4a_{g}' = -a_{g-1}^{(5)} - 4Va_{g-1}''' - 6V'a_{g-1}'' - 2a_{g-1}'V'' + 2a_{g-1}W' + 4a_{g-1}'W\\
0 = -a_g^{(5)} - 4Va_g''' - 6V'a_g'' - 2a_g'V'' + 2a_gW' + 4a_g'W
 \end{array}
\right.  \label{sys}
\end{equation*}
It is clear from the system that the result in \cite{Mironov} can be formulated in the following way. Let $a_1=W/2 + C_1$, where $C_1$ is an arbitrary constant. Define $a_i$ by recursion
\begin{equation}
a_{i+1}= C_{i+1} + \dfrac{1}{4}\int(-a_i^{(5)} - 4Va_i''' - 6V'a_i'' - 2a_i'V'' + 2a_iW' + 4a_i'W)dx
\end{equation}
We see that (3) commutes with an operator of order $4g+2$ and these operators are operators of rank 2 if and only if there exist constants $C_1,...,C_g$ such that $a_{g+1}\equiv const$.
\\
For example,
\begin{equation*}
a_{2}=-\dfrac{1}{8} W^{(4)} - \dfrac{1}{2}VW'' - \dfrac{1}{4}V'W' + \dfrac{3}{8}W^2 + \dfrac{1}{2}WC_1 + 2C_2 ,
\end{equation*}
where $C_2$ is a constant, which appears after integration. In general, $a_i$ contains $i$ constants, i.e., $a_i(x)=a_i(x;C_1,...,C_i)$.\\
\begin{center}
   \textbf{Proof of Theorem 1.}
\end{center}
The operator $L$ has the form
\begin{equation}
L=(\partial^2_x + A_6x^6 + A_2x^2 + \dfrac{A_{-2}}{x^2} + \dfrac{A_{-6}}{x^6})^2 + 16A_6g(g+1)x^4 + B_0.
\end{equation}
Let us prove that we can choose constants $C_1,C_2,...,C_g$ in such a way that  $a_{g+1}=const$. Calculation shows that
\begin{equation*}
a_1= C_1 + \dfrac{B_0}{2} + 8A_6g(g+1)x^4.
\end{equation*}
We see from (6) that $a_{i+1}$ is polynomial in $x$ and linear in $a_i$.
Let us apply (6) to  $x^{4k}$. Assuming that $a_i=x^{4k}$ we get
\begin{equation*}
\begin{gathered}
a_{i+1}= C_{i+1} - 16A_{-6}k(k-1)x^{4k-8} -  k(4k-2)(3+ 4A_{-2} - 16k + 16k^2)x^{4k-4} +\\
+ (B_0 - 16A_2k^2)x^{4k} +  \dfrac{8A_6(g-k)(g+k+1)(2k+1)}{k+1}x^{4k+4}.
\end{gathered}
\end{equation*}
Let us calculate calculate $a_2$ and $a_3$
\begin{equation}
a_2=\widetilde{C}_2 + 4A_6(2C_1 + 3B_0 - 32A_2)g(g+1)x^4 +  96A^2_6g(g+1)(g-1)(g+2)x^8,
\end{equation}
\begin{equation}
\begin{gathered}
a_3 = \widetilde{C}_3 + 4A_6g(1 + g)((16A_2 - B_0)(32A_2 - 3B_0 - 2C_1) +\\ +2(\widetilde{C}_2 - 144A_6(35 + 4A_{-2})(g-1)(g+2))) x^4 -\\
-48A_6^2(160A_2 - 5B_0-2C_1)g(g+1)(g-1)(g+2)x^8 +\\
+1280A_6^3(g - 2)(g - 1)g(g+1)(g+2)(g+3)x^{12},
\end{gathered}
\end{equation}
where $\widetilde{C}_i$ are constants and depend on $C_i$.

If $a_{i}=\widetilde{C_i} +  K_{i}^1x^4 + K^2_{i}x^{8} + ...+K^i_{i}x^{4i}$, then
\begin{equation*}
\begin{gathered}
a_{i+1}=\widetilde{C}_{i+1} + (8A_6g(g+1)\widetilde{C_i}+ (B_0 - 16A_2)K^1_i - 2\cdot 6(35 + 4A_{-2})K^2_i - 16\cdot 2\cdot 3A_{-6} K^3_i)x^4 + \\
+(const\cdot K^1_i + const\cdot K^2_i - const \cdot K^3_{i} - const \cdot K^4_{i})x^{8}  + ...+\\
+\dfrac{8A_6(2i+1)(g-i)(g+i+1)K^i_{i}}{i+1}x^{4i+4}.
\end{gathered}
\end{equation*}
Note that all coefficients, except the leading term, contain constants $C_1,C_2,....$ And $K^{i-1}_{i}$ contains only $C_1$, $K^{i-2}_{i}$ contains only $C_1$ and $C_2$,  $K^{i-3}_{i}$ only $C_1, C_2, C_3$ etc.\\
So, the term $const \cdot x^{4(g+1)}$ in $a_{g+1}$ vanishes because it is multiplied by $const\cdot (g-g)(g+g+1)=0$.
Therefore, we must choose the constants $C_1, C_2,...$ to vanish $K_{g+1}^m=0$ for all $m$. We always can do it because the leading term in $a_{g+1}$ equals $const \cdot x^{4g}$ and depends only on $C_1$, penultimate on $C_1$ and $C_2$ etc.\\
\\
\textbf{Theorem 1 is proved.}\\
\begin{center}
   \textbf{Proof of Theorem 2.}
\end{center}
The operator $L$ has the form
\begin{equation}
L=(\partial^2_x + A_4x^4 + A_2x^2 + A_0 + \dfrac{A_{-2}}{x^2})^2 + 4g(g+1)A_4x^2.
\end{equation}
As in the previous proof, we want to find constants $C_1,C_2,...,C_g$ such that  $a_{g+1}=const$. Calculation shows that
\begin{equation*}
a_1= C_1 + \dfrac{B_0}{2} + 2A_4g(g+1)x^2 .
\end{equation*}
We mentioned before that $a_{i+1}$ is polynomial in $x$ and linear in $a_i$.
Assume that $a_i=x^{2k}$, then from (6)  we have
\begin{equation}
\begin{gathered}
a_{i+1}= C_{i+1} -k(k-1)(3 + 4A_{-2} + 4k(k-2))x^{2k-4} - 2A_0k(2k-1)x^{2k-2} + \\
+  (B_0 - 4A_2k^2)x^{2k} + \dfrac{2A_4(2k+1)(g-k)(g+k+1)x^{2k+2}}{k+1}.
\end{gathered}
\end{equation}
Let us calculate $a_2$ and $a_3$
\begin{equation}
a_2=\widetilde{C_2} + A_4\left(2C_1 + 3B_0 - 8A_2\right)g(g+1)x^2 +6A^2_4g(g+1)(g-1)(g+2)x^4,
\end{equation}
\begin{equation}
\begin{gathered}
a_3= \widetilde{C_3} + A_4g(g+1)\left(\left(4A_2 - B_0\right)\left(8A_2 - 3B_0 - 2C_1\right) + 2\widetilde{C_2} - 72A_0A_4(g-1)(g+2)\right)x^2 -\\
-3A_4^2\left(40A_2 - 5B_0 - 2C_1\right)g(g+1)(g-1)(g+2)x^4  +\\
+  20A^3_4g(g+1)(g-1)(g+2)(g-2)(g+3)x^{6},
\end{gathered}
\end{equation}
where $\widetilde{C}_i$ are constants and depend on $C_i$.\\
\\
If $a_{i}=\widetilde{C}_i +  K_{i}^1x^2 + K^2_{i}x^{4} + ...+K^i_{i}x^{2i}$, then
\begin{equation*}
\begin{gathered}
a_{i+1}=\widetilde{C}_{i+1} + \left(2A_4g(g+1)\widetilde{C}_{i} - (B_0 - 4A_2)K_i^1 -12A_0K_i^2 - 6(15 + 4A_{-2})K^3_i\right)x^2 + \\
+\left(const\cdot K_{i}^1 - const \cdot K^2_{i} - const \cdot K^3_{i} - const \cdot K^4_{i}\right)x^{4}  + .. \\
+ \dfrac{2A_4(2i+1)}{i+1}(g-i)(g+i+1)K^i_{i}x^{2i+2}.
\end{gathered}
\end{equation*}
As in the proof of the previous Theorem all terms except the leading term contain constants $C_1,C_2,...$. And $K^{i-1}_{i}$ contains only $C_1$, $K^{i-2}_{i}$ contains only $C_1,C_2$ and $K^{i-3}_{i}$ only $C_1, C_2, C_3$ etc.\\
The term $const \cdot x^{2(g+1)}$ in $a_{g+1}$ vanishes because it is multiplied by\\ $\dfrac{2A_4(2g+1)}{g+1}(g-g)(g+g+1)=0$. So we must choose constants $C_1, C_2,...$ in such a way that $A_{g+1}^i=0$ for all $i$. We always can do it because the leading term in $a_{g+1}$ depends only on $C_1$, penultimate on $C_1$ and $C_2$ etc.\\
\\
\textbf{Theorem 2 is proved.}\\
\\
\\
\begin{center}
   \textbf{Proof of Theorem 3.}
\end{center}
\textbf{1)} From Theorem 1 we see that the operator
\begin{equation*}
L = \left(\partial_x^2 + A_6x^6 + A_2x^2 + \dfrac{A_{-2}}{x^2} + \dfrac{A_{-6}}{x^6} \right)^2 + 32A_6x^4 + B_0
\end{equation*}
commutes with an operator $M$ of order 6. From (8) we see that if $g=1$, then we must take $C_1 = \dfrac{32A_2 - 3B_0}{2}$. Using  formula (5), where $Q=z + a_1(x)$ we find the spectral curve of operators $L$ and $M$
\begin{equation*}
w^2 = z^3 + \beta_2z^2 + \beta_1z + \beta_0,
\end{equation*}
where
\begin{equation*}
\begin{gathered}
\beta_2 =  (32A_2 - 3B_0,)\\
\beta_1 = \left(256A_2^2 - 64A_2B_0 + 3B_0^2 + 64A_6(4A_{-2} + 3)\right), \\
\beta_0 = -(16A_2 - B_0) \left((16A_2 - B_0)B_0 - 64A_6 (4A_{-2} + 3)\right) - 4096A_6^2A_{-6}.
\end{gathered}
\end{equation*}
So, we must solve the system of equations in $A_6, A_2,A_{-2}, A_{-6}, B_0$
\begin{equation*}
\left\{
 \begin{array}{l}
\beta_2 = q_2\\
\beta_1 =q_1\\
\beta_0 = q_0
 \end{array}
\right.  \label{sys}
\end{equation*}
where $q_2$, $q_1$ and $q_0$ are arbitrary constants. The solution is
\begin{equation*}
\begin{gathered}
A_2 = \dfrac{3B_0 + q_2}{32},\\
A_{-2} = \dfrac{3B_0^2 - 768A_6 + 4q_1 + 2B_0q_2 - q_2^2}{1024A_6},\\
A_{-6} = \dfrac{B_0^3 - 8q_0 + 4B_0q_1 + B_0^2q_2 + 4q_1q_2 - B_0q_2^2 - q_2^3}{32768A_6^2},
\end{gathered}
\end{equation*}
where $A_6 \neq 0, B_0$ are arbitrary constants.\\
\\
\textbf{2)} From Theorem 2 we know that the operator
\begin{equation*}
L = \left(\partial_x^2 + A_4x^4 + A_2x^2 + A_0 + \dfrac{A_{-2}}{x^2}  \right)^2 + 24A_4x^2 + B_0
\end{equation*}
commutes with an operator $M$ of order 10. Again, using formula (5), (12) and (13) we find the spectral curve of operators $L$ and $M$
\begin{equation*}
w^2 = z^5 + \beta_4z^4 + \beta_3z^3 + \beta_2z^2 + \beta_1z + \beta_0,
\end{equation*}
where
\begin{equation*}
\begin{gathered}
\beta_4 = 40A_2 -5B_0, \\
\\
\beta_3 = 2(264A_2^2 + 168A_0A_4 - 80A_2B_0 + 5B_0^2),\\
\\
\beta_2 = 2(1280A_2^3 - 792A_2^2B0 - 504A_0A_4B_0 - 5B_0^3 + 24A_2(156A_0A_4 + 5B_0^2) + 72A_4^2 (13 + 12A_{-2})),\\
\\
\beta_1 = 4096A_2^4 + 27648A_0^2 A_4^2 - 5120A_2^3B_0 - 3744A_4^2B_0 + 1008A_0A_4B_0^2 + 5B_0^4 +\\ +144A_2^2(256A_0A_4 + 11B_0^2) -  3456A_4^2B_0A_{-2} + 32A_2(72A_4^2(17 + 12A_{-2})-468A_0A_4B_0 - 5B_0^3),\\
\\
\beta_0 = 27648A_0^2A_4^2(4A_2 - B_0) + 48A_0A_4(1024A_2^3 - 768A_2^2B_0 + 156A_2B_0^2 - 7B_0^3 +\\
+1728A_4^2(3 + 4A_{-2})) - (16A_2 - B_0)(256A_2^3B_0 - 144A_2^2B_0^2 +\\ + 24A_2(B_0^3 -384A_4^2 ) + B_0(144A_4^2(13 + 12A_{-2})) -B_0^3).
\end{gathered}
\end{equation*}
So, we again must solve the system of equations in $A_4, A_2, A_0, A_{-2}, B_0$
\begin{equation}
\left\{
 \begin{array}{l}
\beta_4 = q_4\\
\beta_3 = q_3 \\
\beta_2 = q_2\\
\beta_1 = q_1\\
\beta_0 = q_0
 \end{array}
\right.
\end{equation}
where $q_4, q_3, q_2, q_1, q_0$ are arbitrary constants. From the first equation we find that \\
\begin{equation*}
B_0 = \dfrac{40A_2 - q_4}{5}.
\end{equation*}
Then we obtain that
\begin{equation*}
\begin{gathered}
A_0 =  \dfrac{560A_2^2 + 5q_3 - 2q_4^2}{1680A_4}\\
A_{-2} = \dfrac{11200A_2^3 - 327600A_4^2 + 175q_2 + 300A_2q_3 - 105q_3q_4 - 120A_2 q_4^2 + 28q_4^3}{302400 A_4^2}\\
A_{2} = \dfrac{6125q_1 - 1500q_3^2 - 2450q_2q_4 + 1935q_3 q_4^2 - 387q_4^4}{56448000A_4^2}
\end{gathered}
\end{equation*}
Hence the last equation of (14) has the form
\begin{equation*}
\begin{gathered}
-\dfrac{2304}{35}A_4^2 (5q_3 - 2q_4^2) + \\
\\
+\dfrac{(-6125q_1 + 1500q_3^2 + 2450q_2q_4 - 1935q_3q_4^2 + 387q_4^4)^2}{86436000000A_4^2}+\\
\\
+\dfrac{625(7q_1 - 12q_3^2)q_4 + 5175q_3q_4^3 - 828q_4^5 + 125q_2(100q_3 - 47q_4^2)}{21875} = \\
\\
=A_4^2h_2 + \dfrac{h_{-2}}{A_4^2} + h_0 = q_0,
\end{gathered}
\end{equation*}
where $h_i$ are coefficients of the equation. We want to find solution $A_4 \neq 0$.

If $q_4=q_3=q_2=q_1=0$ and $q_0=1$, then left side equals zero and right side equals 1. Hence for some $q_i$ we have not solution.

There doesn't exist solution $A_4 \neq 0$ if and only if one of the following system is satisfied
\begin{equation}
\left\{
 \begin{array}{l}
h_2 = 0\\
h_{-2} =  0\\
h_0 \neq q_0
 \end{array}
\right. ,  \quad
\left\{
 \begin{array}{l}
h_2 = 0\\
h_{-2} \neq 0\\
h_0 =  q_0\\
 \end{array}
\right. , \quad
\left\{
 \begin{array}{l}
h_2 \neq 0\\
h_{-2} = 0\\
h_0 = q_0
 \end{array}
\right..
\end{equation}
Now let us consider operator (1) from Theorem 1. If $g=2$, then using (5), (8) and (9), we get that the spectral curve of operators $L$ and $M$ from Theorem 1 has the form
\begin{equation*}
w^2 = z^5 + \gamma_4z^4 + \gamma_3z^3 + \gamma_2z^2 + \gamma_1z + \gamma_0,
\end{equation*}
where
\begin{equation*}
\begin{gathered}
\gamma_4 = 160A_2 - 5B_0, \\
\\
\gamma_3 = 8448A_2^2 - 640A_2B_0 + 10B_0^2 + 192A_6(213 + 28A_{-2}),\\
\\
\gamma_2 = 2(81920A_2^3 - 12672A_2^2B_0 - 5B_0^3 - 288A_6B_0(213 + 28 A_{-2}) +\\ +96A_2(5B_0^2 + 48A_6(359 + 52A_{-2})) + 55296A_6^2A_{-6}),\\
\\
\gamma_1 = 1048576A_2^4 - 327680A_2^3B_0 + 5B_0^4 + 576A_6B_0^2(213 + 28A_{-2}) +\\
+ 2304A_2^2(11B_0^2 + 1024A_6(19 + 4A_{-2})) + 128A_2(55296A_6^2 A_{-6} - 5B_0^3 - 144A_6B_0(359 + 52A_{-2})) +\\
+221184A_6^2(1890 + 496A_{-2} + 32A_{-2}^2 - B_0A_{-6}),\\
\\
\gamma_0 = -1048576A_2^4B_0 - B_0^5 - 192A_6B_0^3(213 + 28A_{-2}) +\\
+32768A_2^3(5B_0^2 + 384A_6(3 + 4A_{-2})) - 768A_2^2B_0(11B_0^2 + 3072A_6(19 + 4A_{-2})) +\\
+84934656A_6^3(35 + 4A_{-2})A_{-6} + 110592A_6^2B_0(-3780 - 992A_{-2} - 64A_{-2}^2 + B_0A_{-6}) +\\
+32A_2(5B_0^4 - 288A_6B_0^2(359 + 52A_{-2}) - 221184A_6^2(105 + 152A_{-2} + 16A_{-2}^2 - B_0A_{-6})).
\end{gathered}
\end{equation*}
So, the system (14) for operator (1) has the form
\begin{equation}
\left\{
 \begin{array}{l}
\gamma_4 = q_4\\
\gamma_3 = q_3 \\
\gamma_2 = q_2\\
\gamma_1 = q_1\\
\gamma_0 = q_0
\end{array}
\right.
\end{equation}
Solving the first,second and the third equations of (16) we get
\begin{equation*}
\begin{gathered}
B_0 = \dfrac{160A_2 - q_4}{5}\\
A_{-2} = \dfrac{8960A_2^2 - 204480A_6 + 5q_3 - 2q_4^2}{26880A_6}\\
A_{-6} = \dfrac{716800A_2^3 + 58982400A_2A_6 + 175q_2 + 1200A_2q_3 - 105q_3q_4 - 480A_2q_4^2 + 28q_4^3}{19353600A_6^2}
\end{gathered}
\end{equation*}
And system (16) is equivalent to the system
\begin{equation*}
\left\{
 \begin{array}{l}
g_1 = 6125q_1\\
g_0 = 153125q_0
 \end{array}
\right.,
\end{equation*}
where
\begin{equation*}
\begin{gathered}
g_1 = 1500q_3^2 -28901376000A_2^2A_6 - 42467328000A_6^2 + 2450q_2q_4 -\\
-1935q_3q_4^2 + 387q_4^4 + 460800A_6(5q_3 - 2q_4^2)\\
\\
g_0 = 87500(6144A_6(q_2 - 2048A_2(7A_2^2 - 72A_6)) + (147456A_2A_6 + q_2)q_3) -\\
-45000(3211264A_2^2A_6 + (768A_6 + q_3)(6144A_6 + q_3))q_4 -\\
-2625(1966080A_2A_6 + 11q_2)q_4^2 + 150(542720A_6 + 177q_3)q_4^3 - 3861q_4^5
\end{gathered}
\end{equation*}
To prove the part 2 of Theorem 3 we must show that the following  systems have solution.
\begin{equation}
\left\{
 \begin{array}{l}
h_2 = 0\\
h_{-2} = 0\\
h_0 \neq q_0\\
g_1 = 6125q_1\\
g_0 = 153125q_0\\
A_6 \neq 0
 \end{array}
\right.  ,
\end{equation}
\begin{equation}
\left\{
 \begin{array}{l}
h_2 = 0\\
h_{-2} \neq 0\\
h_0 = q_0\\
g_1 = 6125q_1\\
g_0 = 153125q_0\\
A_6 \neq 0
 \end{array}
\right. ,
\end{equation}
\begin{equation}
\left\{
\begin{array}{l}
h_2 \neq 0\\
h_{-2} = 0\\
h_0 = q_0\\
g_1 = 6125q_1\\
g_0 = 153125q_0\\
A_6 \neq 0
 \end{array}
\right.
\end{equation}
From system (17) we get
\begin{equation*}
\left\{
 \begin{array}{l}
q_3 = \dfrac{2q_4^2}{5}\\
q_1 = \dfrac{2q_2q_4}{5}  - \dfrac{3q_4^4}{125}\\
q_0 \neq \dfrac{q_2q_4^2}{25} - \dfrac{9q_4^5}{3125}\\
A_6 = -\dfrac{49A_2^2}{72}\\
\\
\dfrac{822083584}{3}A_2^5 - \dfrac{7168(25q_2 - 2q_4^3)}{75}A_2^2 + \dfrac{q_2q_4^2}{25} - \dfrac{9q_4^5}{3125} = q_0\\
A_6 \neq 0
 \end{array}
\right.
\end{equation*}
If equation $\dfrac{822083584}{3}A_2^5 - \dfrac{7168(25q_2 - 2q_4^3)}{75}A_2^2 + \dfrac{q_2q_4^2}{25} - \dfrac{9q_4^5}{3125} = q_0$ has solution $A_2 = 0$, then $q_0 = \dfrac{q_2q_4^2}{25} - \dfrac{9q_4^5}{3125}$ but it contradicts the third equation. So, we obtain that system (17) always has solution.
\\
Now let us consider system (18). We have
\begin{equation*}
\left\{
 \begin{array}{l}
q_3 = \dfrac{2q_4^2}{5}\\
\\
125q_1 - 50q_2q_4 + 3q_4^4 \neq 0\\
\\
q_0 = \dfrac{q_4(625q_1 - 125q_2q_4 + 6q_4^4)}{3125}\\
\\
-4718592A_2^2A_6   = \dfrac{125q_1 - 50q_2q_4 + 3q_4^4}{125} + \dfrac{339738624}{49}A_6^2\\
\\
A_2(\dfrac{3623878656}{7}A_6^2 + 50331648(\dfrac{125q_1 - 50q_2q_4 + 3q_4^4}{4718592 \cdot 125} + \dfrac{339738624}{4718592 \cdot 49}A_6^2)) =\\ \qquad \qquad \qquad \qquad =  - \dfrac{24576}{175}A_6(25q_2 - 2q_4^3)\\
A_6\neq 0
 \end{array}
\right.
\end{equation*}
This system has solution because if $A_6=0$ is solution of

\begin{equation*}
\left\{
 \begin{array}{l}
-4718592A_2^2A_6   = \dfrac{125q_1 - 50q_2q_4 + 3q_4^4}{125} + \dfrac{339738624}{49}A_6^2\\
\\
A_2(\dfrac{3623878656}{7}A_6^2 + 50331648(\dfrac{125q_1 - 50q_2q_4 + 3q_4^4}{4718592 \cdot 125} + \dfrac{339738624}{4718592 \cdot 49}A_6^2)) =\\ \qquad \qquad \qquad \qquad =  - \dfrac{24576}{175}A_6(25q_2 - 2q_4^3)\\
 \end{array}
\right.
\end{equation*}
then $125q_1 - 50q_2q_4 + 3q_4^4 = 0$ but there exists solution $(A_2, A_6)$ for all $q_i$.
So, we obtain that system (18) has solution.

Finally, let us consider system (19). We get
\begin{equation*}
\left\{
 \begin{array}{l}
5q_3 - 2q_4^2 \neq 0\\
q_1 = \dfrac{1500q_3^2 + 2450q_2q_4 - 1935q_3q_4^2 + 387q_4^4}{6125}\\
q_0 = \dfrac{(100q_3 - 33q_4^2) (875q_2 - 450q_3q_4 + 117q_4^3)}{153125}\\
\\
A_6(62720A_2^2 + 92160A_6 - 5q_3 + 2q_4^2) = 0\\
\\
A_6(92160A_6(1120A_2 - 3q_4) -10035200A_2^3 + 700q_2 -\\
\quad \quad \quad \quad \quad \quad \quad - 188160A_2^2q_4-405q_3q_4 + 106q_4^3+3360A_2(5q_3 - 2q_4^2)) = 0\\
A_6\neq 0.
 \end{array}
\right.
\end{equation*}
Then
\begin{equation*}
\left\{
 \begin{array}{l}
5q_3 - 2q_4^2 \neq 0\\
q_1 = \dfrac{1500q_3^2 + 2450q_2q_4 - 1935q_3q_4^2 + 387q_4^4}{6125}\\
\\
q_0 = \dfrac{(100q_3 - 33q_4^2) (875q_2 - 450q_3q_4 + 117q_4^3)}{153125}\\
\\
A_6 = \dfrac{-62720A_2^2 + 5q_3 - 2q_4^2}{92160}\\
\\
(62720A_2^2 - 5q_3 + 2q_4^2)(2867200A_2^3 - 160A_2(5q_3 - 2q_4^2) - 25q_2 + 15q_3q_4 - 4q_4^3) =0\\
A_6 \neq 0
 \end{array}
\right.
\end{equation*}
This system has solution. So, the part 2 of Theorem 3 is proved.
\\
\\
\textbf{3)}. From Theorem 2 we know that operator
\begin{equation*}
L = \left(\partial_x^2 + ax^4 + bx^2 + c + \dfrac{d}{x^2}  \right)^2 + 48ax^2 + e, \quad a\neq 0
\end{equation*}
commutes with an operator $M$ of order 14. Using formula (5) we find the spectral curve of operators $L$ and $M$
\begin{equation*}
w^2 = z^7 + \beta_6z^6 + \beta_5z^5 + \beta_4z^4 + \beta_3z^3 + \beta_2z^2 + \beta_1z + \beta_0,
\end{equation*}
where
\begin{equation*}
\begin{gathered}
\beta_6 = -7(e - 16b), \\
\\
\beta_5 = 21 (224 b^2 + 96 a c - 32 b e + e^2),\\
\\
\beta_4 = (92416 b^3 + 3168 a^2 (37 + 12 d) + 288 a c (516 b - 35 e) - 23520 b^2 e + 1680 b e^2 - 35 e^3),\\
\\
\beta_3 = 872704 b^4 - 369664 b^3 e + 47040 b^2 e^2 - 2240 b e^3 + 35 e^4 +\\ 1152 a^2 (930 c^2 + 8 b (749 + 264 d) - 11 (37 + 12 d) e) +  576 a c (6312 b^2 - 1032 b e + 35 e^2),\\
\\
\beta_2 = 3 (1204224 b^5 + 115200 a^3 c (341 + 108 d) - 872704 b^4 e + 184832 b^3 e^2 -\\ 15680 b^2 e^3 + 560 b e^4 - 7 e^5 +\\ 192 a c (62048 b^3 - 18936 b^2 e + 1548 b e^2 - 35 e^3) + 192 a^2 (8 b^2 (24107 + 9972 d) +\\ 48 b (1410 c^2 - 749 e - 264 d e) + 3 e (-1860 c^2 + 407 e + 132 d e))).
\end{gathered}
\end{equation*}
We must solve the system of equations in $a, b, c, d, e$
\begin{equation}
\left\{
 \begin{array}{l}
\beta_6 = q_6\\
\beta_5 = q_5\\
\beta_4 =q_4\\
\beta_3 = q_3\\
\beta_2 = q_2
\end{array}
\right.
\end{equation}
where $q_6,q_5,q_4$, $q_3$ and $q_2$ are arbitrary constants. Solving the first, second, third and fourth equations of system (20) we obtain
\begin{equation}
\left\{
 \begin{array}{l}
e  = \dfrac{112b - q_6}{7}\\
c = \dfrac{4704b^2 + 7q_5 - 3q_6^2}{14112a}\\
d = \dfrac{68992b^3 - 5743584 a^2  + 49q_4 + 308bq_5 - 35q_5q_6 - 132bq_6^2 + 10q_6^3}{1862784a^2}\\
b=  \dfrac{7595q_5^2 - 28812q_3 + 16464q_4q_6 - 12390q_5q_6^2 + 2655q_6^4}{17259540480 a^2}\\
h_2a^2 + \dfrac{h_{-2}}{a^2} + h_0 = q_2
\end{array}
\right.
\end{equation}
where
\begin{equation*}
\begin{gathered}
h_2 = \dfrac{9600}{49}(7q_5 - 3q_6^2)\\
h_{-2} = \dfrac{(28812q_3 - 7595q_5^2 - 16464q_4q_6 + 12390q_5q_6^2 - 2655q_6^4)^2}{323233222201344}\\
h_0 = \dfrac{22638q_3q_6 - 18375q_5^2q_6 + 14665q_5q_6^3 - 2514q_6^5 +1029q_4(25q_5 - 17q_6^2)}{52822}
\end{gathered}
\end{equation*}
Arguing as before we see that we have not solution $a\neq 0$ of (21) if and only if one of the following system is satisfied.
\begin{equation}
\left\{
 \begin{array}{l}
h_2 = 0\\
h_{-2} =  0\\
h_0 \neq q_0
 \end{array}
\right. ,  \quad
\left\{
 \begin{array}{l}
h_2 = 0\\
h_{-2} \neq 0\\
h_0 =  q_0\\
 \end{array}
\right. , \quad
\left\{
 \begin{array}{l}
h_2 \neq 0\\
h_{-2} = 0\\
h_0 = q_0
 \end{array}
\right..
\end{equation}
Let us consider the operator
\begin{equation*}
L = \left(\partial_x^2 + ax^6 + bx^2  + \dfrac{c}{x^2} + \dfrac{d}{x^6} \right)^2 + 192ax^4 + e, \quad a\neq 0
\end{equation*}
If $g=3$, then the spectral curve of operators $L$ and $M$ from Theorem 1 has the form
\begin{equation*}
w^2 = z^7 + \gamma_6z^6 + \gamma_4z^4 + \gamma_3z^3 + \gamma_2z^2 + \gamma_1z + \gamma_0,
\end{equation*}
where
\begin{equation*}
\begin{gathered}
\gamma_6 = 7 (64b - e), \\
\\
\gamma_5 = 3 (384a(501 + 28c) + 7(3584b^2 - 128be + e^2)),\\
\end{gathered}
\end{equation*}
\begin{equation*}
\begin{gathered}
\gamma_4 = 5914624b^3 + 2433024a^2d - 376320b^2e + 6720be^2 - 35e^3 +\\
+ 1152a(16b(8227 + 516c) - 5e(501 + 28c)),
\end{gathered}
\end{equation*}
\begin{equation*}
\begin{gathered}
\gamma_3 = 223412224b^4 - 23658496b^3e + 752640b^2e^2 - 8960be^3 + 35e^4 +\\
+ 110592a^2(755475 + 87560c + 2480c^2 + 5632bd -88de) +\\
+ 2304a(128b^2(41567 + 3156c) - 32be(8227 + 516c) + 5e^2(501 + 28c)),\\
\\
\gamma_2 = 3(1233125376b^5 + 353894400a^3d(635 + 36c) - 223412224b^4e +\\
 +11829248b^3e^2 - 250880b^2e^3 + 2240be^4 - 7e^5 + 768a(2048b^3(77977 + 7756c) -\\
  -384b^2e(41567 + 3156c) + 48be^2(8227 + 516c) - 5e^3(501 + 28c)) +\\
   36864 a^2 (425472 b^2 d + 32b(1840425 + 311960 c + 11280 c^2 - 528de) -\\
      3e(755475 + 87560c + 2480 c^2 - 44 de))).
\end{gathered}
\end{equation*}
Solving the first, second and the third equations we get
\begin{equation}
\left\{
 \begin{array}{l}
e  = \dfrac{448b - q_6}{7}\\
c = \dfrac{75264b^2 - 4040064a + 7q_5 - 3q_6^2}{225792a}\\
\\
d = \dfrac{908328960ab + 4415488b^3 + 49q_5 + 1232bq_5 - 35q_5q_6 - 528bq_6^2 + 10q_6^3}{119218176 a^2}\\
g_3 = q_3\\
g_2 = q_2
\end{array}
\right.
\end{equation}
where
\begin{equation*}
\begin{gathered}
g_3 = 306708480 a b^2 + \frac{4q_4q_6}{7} +\\+ \dfrac{5(1519q_5^2 - 10987147100160a^2  - 2478q_5q_6^2 + 531q_6^4 - 3354624a(7q_5 - 3q_6^2))}{28812},\\
\\
g_2 = 1258291200ab^3 + \dfrac{920125440}{7}ab^2q_6 - \frac{2457600}{49}ab(3354624a + 7q_5 - 3q_6^2) - \\
-\dfrac{604293090508800a^2q_6 + 173705q_5^2q_6 - 69020q_5q_6^3 + 5991q_6^5}{739508}+\\
+\dfrac{2058q_4(175q_5 - 31q_6^2) - 430080a(6860q_4 - 1897q_5q_6 + 113q_6^3)}{739508}
\end{gathered}
\end{equation*}
As in the previous part, to prove the part 3 of Theorem 3 we must show that the following  systems have solution.
\begin{equation}
\left\{
 \begin{array}{l}
h_2 = 0\\
h_{-2} = 0\\
h_0 \neq q_2\\
g_3 = q_3\\
g_2 = q_2\\
a \neq 0
 \end{array}
\right.  ,
\end{equation}
\begin{equation}
\left\{
 \begin{array}{l}
h_2 = 0\\
h_{-2} \neq 0\\
h_0 = q_2\\
g_3 = q_3\\
g_2 = q_2\\
a \neq 0
 \end{array}
\right. ,
\end{equation}
\begin{equation}
\left\{
\begin{array}{l}
h_2 \neq 0\\
h_{-2} = 0\\
h_0 = q_2\\
g_3 = q_3\\
g_2 = q_2\\
A_6 \neq 0
 \end{array}
\right.
\end{equation}
From system (24) we obtain
\begin{equation*}
\left\{
 \begin{array}{l}
q_5 = \dfrac{3q_6^2}{7}\\
q_3 = \dfrac{196q_4q_6 - 15q_6^4}{343}\\
q_2 \neq \dfrac{3(98q_4q_6^2 - 9q_6^5)}{2401} \\
a = \dfrac{637b^2}{3960}\\
\\
-\dfrac{502293069824}{121}b^5  - \dfrac{33280(49 q4 - 5 q6^3)}{2541}b^2 + \dfrac{3(98q_4q_6^2 - 9q_6^5)}{2401} = q_2\\
a \neq 0
 \end{array}
\right.
\end{equation*}
We see that this system has solution.

From system (25) we get
\begin{equation*}
\left\{
 \begin{array}{l}
q_5 = \dfrac{3q_6^2}{7}\\
\\
343q_3 - 196q_4q_6 + 15q_6^4 \neq 0\\
\\
q_2 = \dfrac{3q_6(343q_3 - 98q_4q_6 + 6q_6^4)}{2401}\\
\\
306708480ab^2 = q_3 + \dfrac{93428121600}{49}a^2 - \dfrac{4q_4q_6}{7} + \dfrac{15q_6^4}{343}\\
\\
-\dfrac{943718400}{343}a(20384b + 99q_6) + \dfrac{20480a(77271040b^3 - 245q_4 + 8072064b^2q_6 + 25q_6^3)}{3773} -\\
\\
\qquad \qquad \qquad -\dfrac{q_6(343q_3 - 196q_4q_6 + 15q_6^4)}{2401} = 0\\
a \neq 0
 \end{array}
\right.
\end{equation*}
If $a=0$ is root of fourth equation, then $343q_3 - 196q_4q_6 + 15q_6^4 = 0$ but it contradicts the second equation. But the fourth and fifth equations have solutions (a,b) for all $q_i$.
\\
System (26) has the form
\begin{equation*}
\left\{
 \begin{array}{l}
7q_5 - 3q_6^2 \neq 0\\
\\
q_3 = \dfrac{7595q_5^2 + 16464q_4q_6 - 12390q_5q_6^2 + 2655q_6^4}{28812}\\
\\
q_2 = \dfrac{-173705q_5^2q_6 + 69020q_5q_6^3 - 5991q_6^5 + 2058q_4(175q_5 - 31q_6^2)}{739508}
\\
a = \dfrac{13 (526848b^2 - 7q_5 + 3q_6^2)}{42577920}\\
\\
(526848 b^2 - 7 q5 + 3 q6^2)(49(97026048b^3 + 15q_4 - 32bq_5) - 525q_5q_6 + 672bq_6^2 + 150q_6^3) = 0\\
a\neq 0.
 \end{array}
\right.
\end{equation*}
This system has solution.\\
\\
\textbf{Theorem 3 is proved.}

\end{document}